\documentclass[a4paper,conference]{IEEEtran}
\usepackage[utf8]{inputenc}
\usepackage{algorithm}
\usepackage{algpseudocode}
\usepackage{url}
\usepackage{graphics}
\usepackage{graphicx}
\usepackage{color}
\usepackage{multirow}
\usepackage{dblfloatfix}

\def\ermloc{(x_m^\epsilon, y_m^\epsilon, z_m^\epsilon)}
\def\exmloc{(x_m, y_m, z_m)}
\def\opmloc{(x_m', y_m', z_m')}
\def\rir{h}
\def\exsloc{(x_s, y_s, z_s)}
\def\x{u}
\def\o{y}
\def\noise{n}
\def\asr{\mbox{\tt gp-asr}}
\def\gpASR{\asr}
\def\ed{D}
\def\cer{\mbox{\tt cer}}
\def\CER{\cer}
\def\r{\gamma}
\def\samloc{(x_m^i, y_m^i,z_m^i)}
\def\samlocs{i}
\def\opmloc{(x_m^*, y_m^*,z_m^*)}
\def\Generate{{\sc Gen}}
\def\SS{{\sc ss}}
\def\paper{paper}

\title{Computing Optimal Location of Microphone for Improved Speech Recognition}
\author{\IEEEauthorblockN{Karan Nathwani}
\IEEEauthorblockA{Department of Electrical Engg\\
IIT Jammu, India\\
Email: karan.nathwani@iitjammu.ac.in}
\and
\IEEEauthorblockN{Bhavya Dixit}
\IEEEauthorblockA{Department of Electrical Engg\\
IIT Jammu, India\\
Email: 2020pee1035@iitjammu.ac.in}
\and
\IEEEauthorblockN{Sunil Kumar Kopparapu}
\IEEEauthorblockA{TCS Research\\
Tata Consultancy Services Limited, India\\
Email: sunilkumar.kopparapu@tcs.com}}

\begin{document}
\maketitle
\IEEEpeerreviewmaketitle
\begin{abstract}
It  was  shown  in  our earlier work 
that the measurement error in the microphone position  affected  the  room  impulse  response (RIR) which in turn affected the single-channel close microphone \cite{Raikar2020} and multi-channel distant microphone  \cite{Nathwani2021} speech recognition.    
In this paper, as an extension, we systematically study to identify the {\em optimal location} of the microphone, given an approximate and hence erroneous location of the microphone in 3D space. 
The primary idea is to use Monte-Carlo technique to generate 
a large number of random microphone positions around the 
erroneous microphone position and select the 
microphone position that results in the best performance of a general purpose automatic speech recognition (\gpASR). We experiment with clean and noisy speech and show that the optimal location of the microphone is unique and is affected by noise. 
\end{abstract}

\section{Introduction}
\label{sec:introduction}

In the last few decades, the paradigm has been shifted to address the 
practical scenario that degrades the performance of a general purpose automatic speech recognition (\asr) system.
One of the common scenarios is the 
location of microphones that get disturbed during routine maintenance or due to human interventions \cite{muthukumarasamy2009impact,sachar2002position}. As a result of the disturbance in the  microphone location (and hence erroneous), the room impulse response (RIR) is impacted; the correct (or actual) location of the microphone is desired to obtain the exact RIR which in turn has an impact on the performance of a \asr. 

Smart Speakers (\SS) have become a very popular commercial gadget which gives the flexibility to transact using voice based queries.
The user-experience of a \SS\ depends on its ability to {\em understand} the spoken query which significantly  depends on the performance of a 
\gpASR\ 
engine, an important part of \SS\ pipeline.
Consequently, the correct position of the microphone 
mounted on a \SS\ is crucial in computing the 
RIR
\cite{szoke2019building}. It is well known that 
the quality \cite{duangpummet2022blind,chen2021noising,wei2021exploring}  and intelligibility \cite{nathwani2017speech,nathwani2016formant,biswas2021transfer} of the speech signal at the microphone is affected by RIR. 
This results in degradation of the performance of the \gpASR \cite{naylor2010speech,karpagavalli2016review}. In a nutshell, an error in 
the exact location of the microphone, 
leads to an error in computing the RIR which inturn leads to poor performance of \gpASR\ 
\cite{Raikar2020,Nathwani2021}. 

In literature, some attempts have been made in estimating the RIR's in single channel scenario by assuming the 
{actual}
microphone position is known {\em a priori}. 
A maximum a posteriori (MAP) estimation of RIR has been attempted by \cite{florencio2015maximum} for a single microphone scenario. 
	The inverse filtering of RIR has also been carried out in \cite{mosayyebpour2010single} which maximizes the skewness of linear predictive {\color{black}{(LP)}} 
	residual. This method exploits the adaptive gradient descent method for the computation of inverse filtering of RIR. 
	Subsequently, 
	\cite{padaki2013single} 
	tried to utilize the LP residual cepstrum for the de-reverberation task.
	More recently, \cite{raikar2018single}
	proposed a joint noise cancellation and  de-reverberation approach using deep priors by focusing on the noisy estimate of RIR. 
	The sparsely estimated RIR study in \cite{mohanan2017speech} has also been used as a regularization parameter in non-negative matrix factorization {\color{black}{(NMF)}} framework for speech de-reverberation. 
	 The above attempts assume the availability of exact location of microphone position while  computation RIR. However finding the optimal position of the microphone after displacement from its original known location has been less explored.

In related work, a few studies have concentrated on identifying 
the optimal microphone position assuming 
the availability of knowledge of inter intensity difference and inter time difference 
from 
different microphones \cite{huang2003class, juhlin2021optimal,
jiang2019group}. An analytical model of location errors using delay and sum beam-forming is studied in \cite{muthukumarasamy2009impact}. The delay and sum beamforming is replaced by minimum variance distortionless response (MVDR) beamforming in \cite{zhang2017microphone,nathwani2014speech} for noise reduction tasks using an optimal subset of microphones array. 
While a microphone array configurations are beneficial 
compared to single microphone in more practical scenario 
\cite{nathwani2013multi,1197303} 
the single microphone case is more challenging because of the 
unavailability of cues like, direction-of-arrival which makes identifying the location of microphone 
non-trivial \cite{sachar2004microphone, sachar2002position}. 
To the best of the author's knowledge, no attempt has been made in finding the optimal microphone location when a single microphone scenario is considered. 

	%


In this paper, we focus on identifying the {\em exact location} of the (displaced) microphone, given an approximate or erroneous location of the microphone. The primary idea is to use Monte-Carlo technique to generate a distribution of random microphone positions around the known but erroneous microphone position and then identify that microphone position that results in the best performance of the \gpASR. 
The rest of the \paper\ is organized as follows, in Section \ref{sec:problem} we formulate the problem and conduct extensive experiments in Section \ref{sec:experiments} and analyze the findings. We conclude in Section \ref{sec:conclusions}.

\section{Problem Formulation}
\label{sec:problem}
Let us say $\exmloc$ is the exact location of the microphone, however due to an error in measurement of the location of the microphone, we are given the location of the microphone as being, say $\ermloc$. If we were to use  $\ermloc$ to compute the room impulse response (RIR), say $\rir^\epsilon(t)$, then there is a degradation in both the performance of an automatic speech recognition (\gpASR) engine and also the intelligibility of speech as elaborated in \cite{Raikar2020}. In this \paper, we try to find the exact (or optimal) location of the microphone, $\opmloc$ given the inexact location of the microphone, namely $\ermloc$. 

\begin{quote} \em 
Given the inexact location of the microphone, find the exact or optimal location of the microphone,
which results in the best performance of a \gpASR\ or equivalently results in the least character error rate (\cer). 
\end{quote}

Let $\exmloc$ and $\exsloc$ be the location of the microphone and the speaker respectively,  $\x(t)$ be the utterance spoken by the speaker, and $\x_{txt}$ be the text that is spoken to produce $\x(t)$. The output of the microphone is given by \begin{equation} 
\o(t) = \rir(t) * \x(t) + \noise(t)
\label{eq:yequalhxplusn}
\end{equation}
where $\rir(t)$ is the RIR which depends on the location of the speaker and the microphone position (see \cite{Raikar2020}) and $\noise(t)$ is the ambient environmental noise. Let $\asr$ be an ASR engine which converts a given utterance into text, namely,\begin{equation}
    \o_{txt} = \asr(\o(t)).
\end{equation} Now we compute the performance of the ASR by computing the edit distance \cite{10.1145/375360.375365} $\ed$ between the two text string, namely, $\x_{txt}$ (ground truth) and $\o_{txt}$ (\asr\ output), namely, \begin{equation} \cer(\exmloc) =\ed(\x_{txt}, \o_{txt}).\end{equation}
 Let us assume that the actual position $\exmloc$ is within a radial distance $\r$ from the measured location $\ermloc$, namely, 
\begin{equation}
\sqrt{(x_m-x_m^\epsilon)^2 +
(y_m-y_m^\epsilon)^2 +
(z_m-z_m^\epsilon)^2} \le \r
\end{equation}
\begin{algorithm}[!ht]
\caption{Computing Optimal position of microphone}
\begin{algorithmic}[1]
\State Given $N$, $\r$, $\ermloc$, $\x(t)$, $\x_{txt}$, $\noise(t)$, $\asr$, $\cer$

\State Output: $\opmloc$

\For{$i\leq N$}
\State $\samloc$ = \Call{\Generate}{$\r,\ermloc$} 

\State Compute $\rir^i(t)$ \Comment{See \cite{Raikar2020}}

\State $\o^i(t) = \rir^i(t) * \x(t) + \noise(t)$

\State $\o^i_{txt} = \asr(\o^i(t))$

\State $\cer(\samlocs) = \ed(\x_{txt}, \o^i_{txt})$

\EndFor

\State $\opmloc = \min \cer(\samlocs)$

\State

\Function{\Generate}{$\r,\ermloc$}\Comment{Monte Carlo Method}

\State $\samloc = \ermloc + {\mathbf N}(0,\r^2)$ 

\Comment{${\mathbf N}(0,\r^2) = (N(0,\r^2),N(0,\r^2),N(0,\r^2))$}

\Comment{$N(0,\r^2)$ is 0 mean Gaussian with variance $\r^2$}

\State{$D = \sqrt{(x_m^i-x_m^\epsilon)^2 +
(y_m^i-y_m^\epsilon)^2 +
(z_m^i-z_m^\epsilon)^2}$}
\While{$D \le \r$}

\State $\samloc = \ermloc + {\mathbf N}(0,\r^2)$

\EndWhile

\Return $\samloc$
\EndFunction
\end{algorithmic}
\label{algo:optimal_mpos}
\end{algorithm}

We use Monte Carlo method \cite{enwiki:1043540175} to generate $\samloc$ such that it is within a radius of $\r$ from the known inexact microphone location $\ermloc$. 
We use $\samloc$ to compute RIR, $\rir^i(t)$ and obtain $\o^i(t) = \rir^i(t) * \x(t) + \noise(t)$. The output of the ASR $\o^i_{txt} = \asr(\o^i(t))$ is used to compute  $\cer(\samlocs) = \ed(\x_{txt}, \o^i_{txt})$. The $\samloc$ value for which  $\cer(\samlocs)$ is minimum determines the optimal location of the microphone. Mathematically,
\begin{equation}
\opmloc = \min_{\samloc} \cer(\samlocs)
\label{eq:mincer}
\end{equation}
Note that $\samloc$ is constrained to be radially within $\r$ distance of $\ermloc$ (Details in Algorithm \ref{algo:optimal_mpos}).

\begin{figure}[!b]
    \centering
    \includegraphics[width=0.45\textwidth]{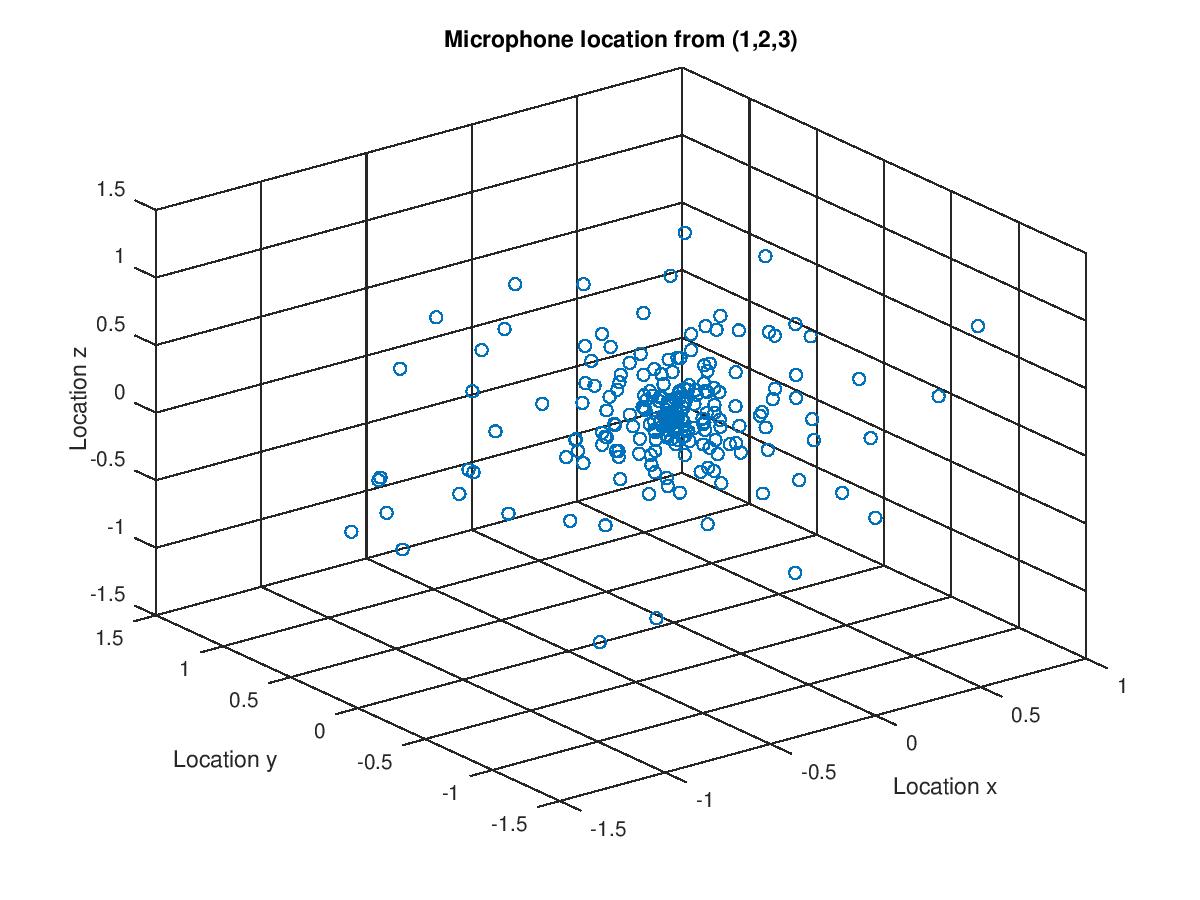}
    \caption{Microphone locations (using $\r=0.02,0.1,0.5,1.0,1.5$) used in our simulations. A total of $500$ locations.}
    \label{fig:mic_positions}
\end{figure}

\begin{figure}
    \centering
    \includegraphics[width=0.45\textwidth]{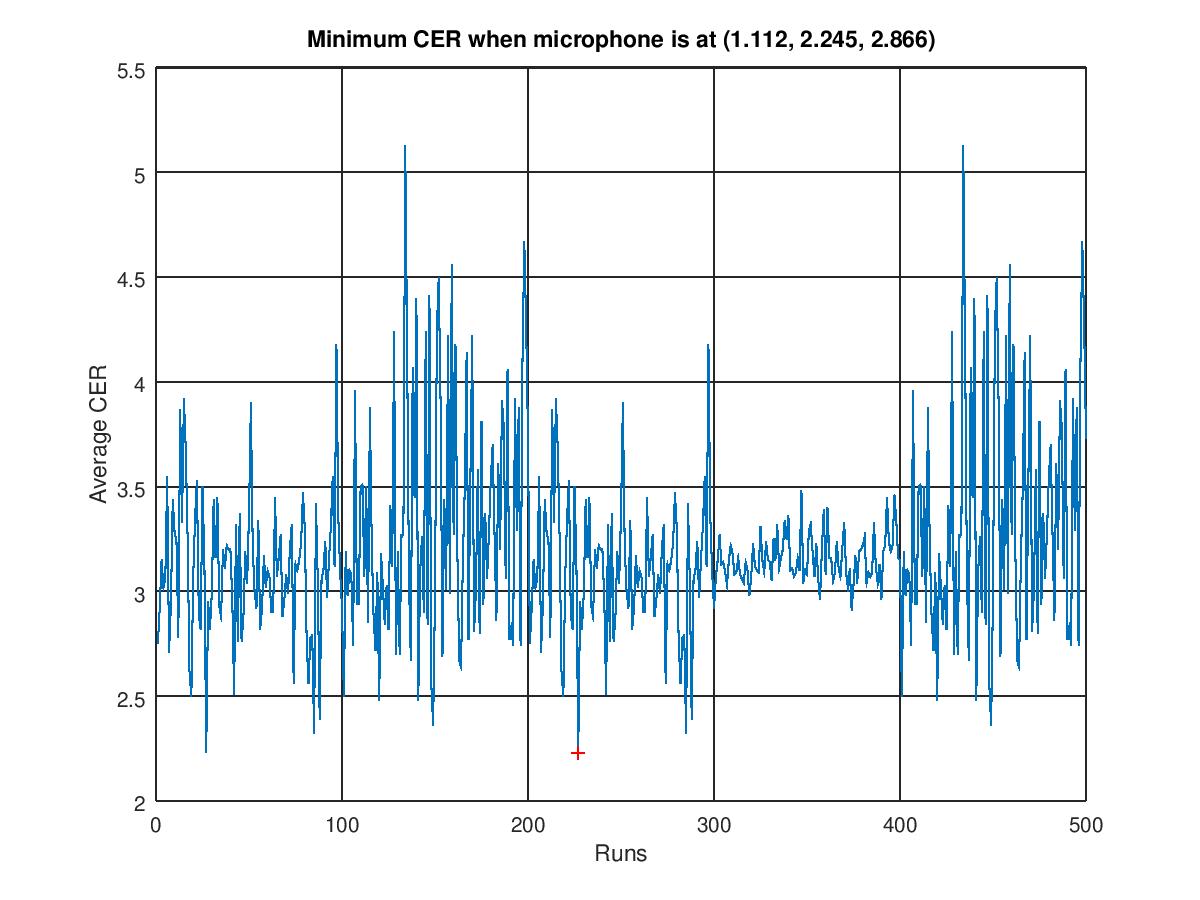}
    
    (a)
    
     \includegraphics[width=0.45\textwidth]{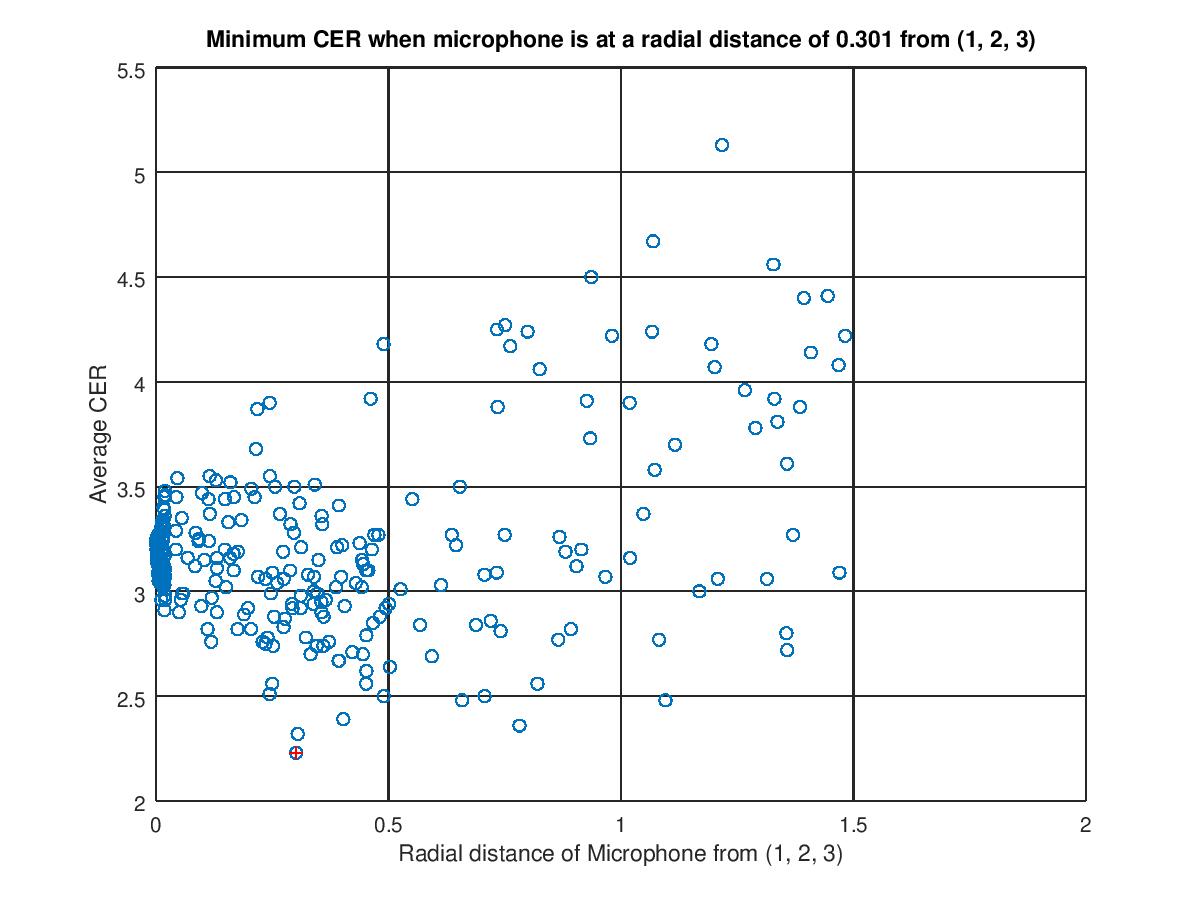}
     
     (b)
     
    \caption{Average \cer\ over $100$ utterances for $500$ different microphone positions. The optimal microphone location marked in red produces the minimum average \cer\ ($=2.23$).}
    \label{fig:av_cer_runs}
\end{figure}

\section{Experimental Results}
\label{sec:experiments}

\subsection{Experimental Setup}

We used 
the office room ({\tt L212}) dimensions 	$(7.5 \times 4.6 \times 3.1) m^3$ mentioned in \cite{szoke2019building} for all our experiments. For each $\r = 0.02, 0.1, 0.5, 1.0, 1.5$ we generated $100$ microphone locations, so in all  we had in all $500$ different locations of the microphone (Figure \ref{fig:mic_positions} shows the location of the microphone from the inexact location of the microphone at $(1, 2, 3)$) which we used in our experiments. For each of the $500$ microphone positions, we took $100$ utterances ($u(t)$) from LibreSpeech-dev \cite{librespeech} picked at random and for each utterance we computed ($y(t)$) using the microphone position to compute the $\rir$. We then converted $y(t)$ into alphabets using a transformer based speech to alphabet recognition engine \cite{wav2vec-transformer} and finally computed the \cer\ for each utterance. We averaged the \cer\ across the $100$ utterances. 

\subsection{Optimal Microphone Position in the absence of Noise}
Figure \ref{fig:av_cer_runs}(a) shows the average \cer\ for each of the $500$ microphone positions. The least average \cer\ is obtained for the microphone position, $(1.11, 2.24, 2.86)$ which is marked in red in Figures \ref{fig:mic_positions}, 
and  \ref{fig:av_cer_runs}.
%
Figure \ref{fig:av_cer_runs}(b) shows the radial distance between the microphone location from $(1, 2, 3)$ and the corresponding average \cer\ for that position over $100$ utterances. The optimal microphone position, producing the least average \cer\ ($=2.23$) is shown in red and is at a radial distance of $0.301$ from $(1, 2, 3)$. Note that there is no consistency between the average \cer\ and radial distance of the microphone from $(1, 2, 3)$ and this is one of the reasons why a {\em closed form solution to obtain the optimal position of microphone locations eludes us}. Note that while there is an observable increase in average \cer\ with increasing radial distance of the microphone but for a given average \cer\ one can find the radial distance of the microphone can range from $0$ to $1.5$. 


\begin{figure}
    \centering
\includegraphics[width=0.45\textwidth]{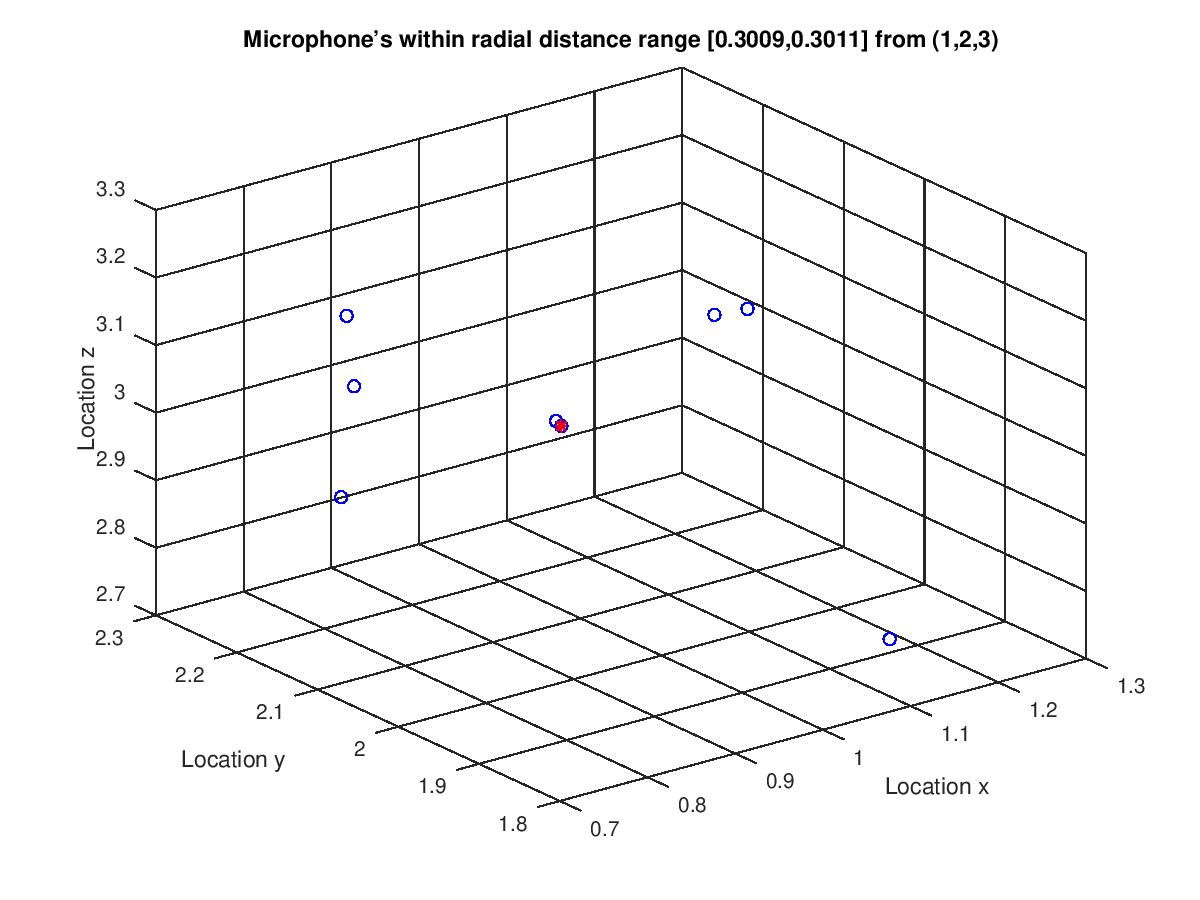}

(a)

\includegraphics[width=0.45\textwidth]{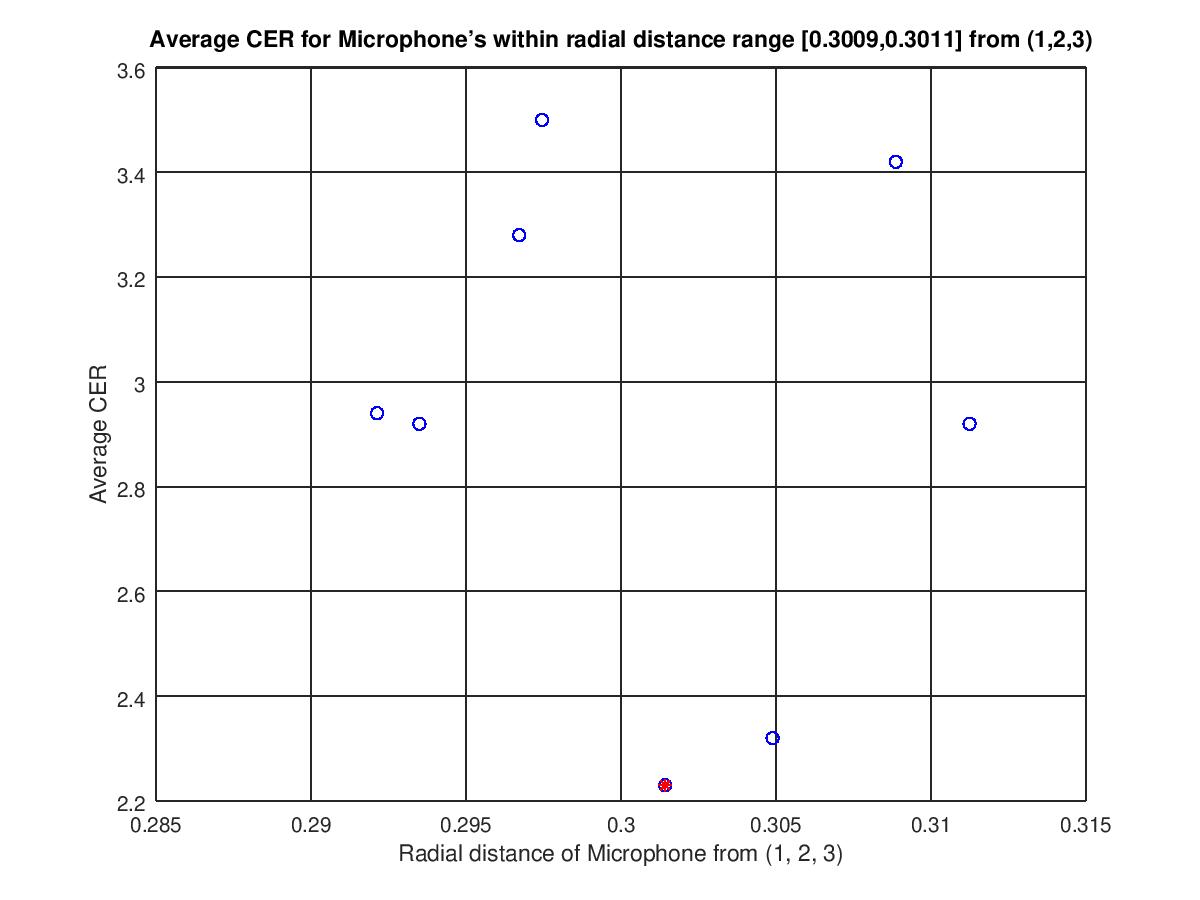}

(b)

\caption{(a) Microphone positions within a distance of  $\pm 0.0001$ from the optimal microphone position (in red). (b) Average \cer\ associated with the microphone positions in (a).}
    \label{fig:mic_pos_range}
\end{figure}
We wanted to observe if all microphone positions which were radially at the same distance as the optimal microphone position resulted in similar average \cer.  Since there were no microphone location exactly with $\r=0.301$, we identified all microphone locations which were within $\pm 0.0001$ from the optimal microphone position. Figure  \ref{fig:mic_pos_range}(a) shows the location of all the microphones (among the $500$ microphone locations) which were within $\pm 0.0001$ from the optimal location of the microphone (radial distance $0.301$) marked in red. As seen in Figure \ref{fig:mic_pos_range}, even if the microphones are close (bounded within $\pm 0.0001$) to the optimal location of the microphone, the average \cer\ shows no  such bound (see Figure \ref{fig:mic_pos_range}(b)). This again establishes the fact that there is no association between the microphone location and the average \cer\ to warrant an analytical solution to find the optimal location of the microphone. 

\begin{figure}
    \centering
    \includegraphics[width=0.45\textwidth]{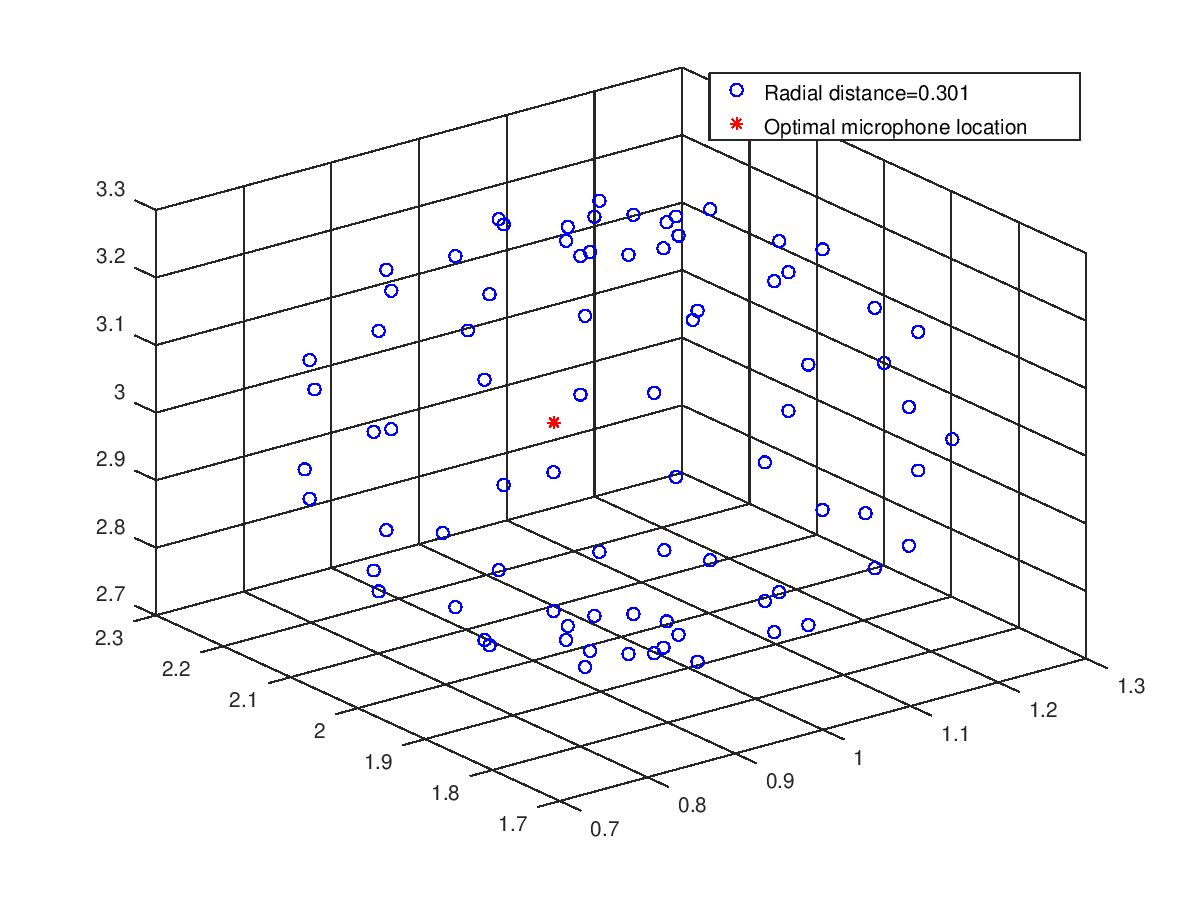}
    \caption{Microphone position at the radial distance of $0.301$ from $(1, 2, 3)$. The optimal location of the microphone is marked with a red star.}
    \label{fig:rad3}
\end{figure}

To ascertain more concretely if this was true, we choose an additional  $100$ microphone locations which were {\em exactly} at a radial distance of $0.301$ (see Figure \ref{fig:rad3}). The minimum average \cer\ for $100$ utterances for each of the $100$ microphone locations was $2.37$ (higher than that $2.23$ obtained for the identified optimal location of microphone). This shows that it is not sufficient for the microphones to be a particular radial distance but the {\em actual} location in the 3D space is crucial.

%
\subsection{Optimal Microphone Position in the presence of Noise}
In the next set of experiments, we introduced two kinds of noise, namely, (a) additive white Gaussian noise and (b) real {\em home} environment noise \cite{w3xm-jn45-20} in (\ref{eq:yequalhxplusn}) such that the resulting signal had an SNR of $0, 5, 10, 15$ dB. 
\begin{figure}
    \centering
    \includegraphics[width=0.5\textwidth]{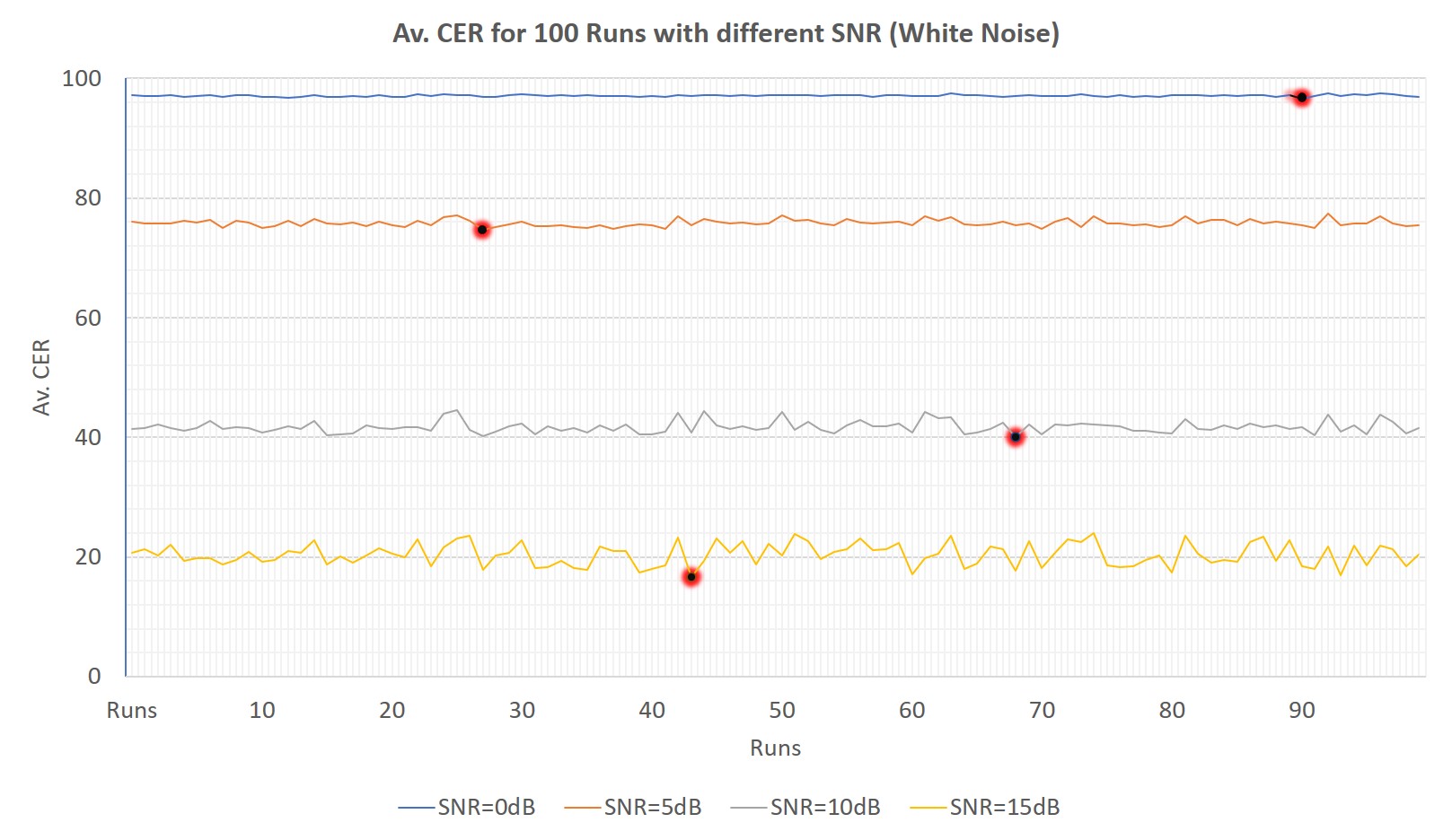}
    \caption{Average \CER\ for $100$ runs for varying SNRs in Additive White Gaussian Noise (AWGN).}
    \label{fig:noise_cer_runs1}
\end{figure}

\begin{figure}
    \centering
    \includegraphics[width=0.5\textwidth]{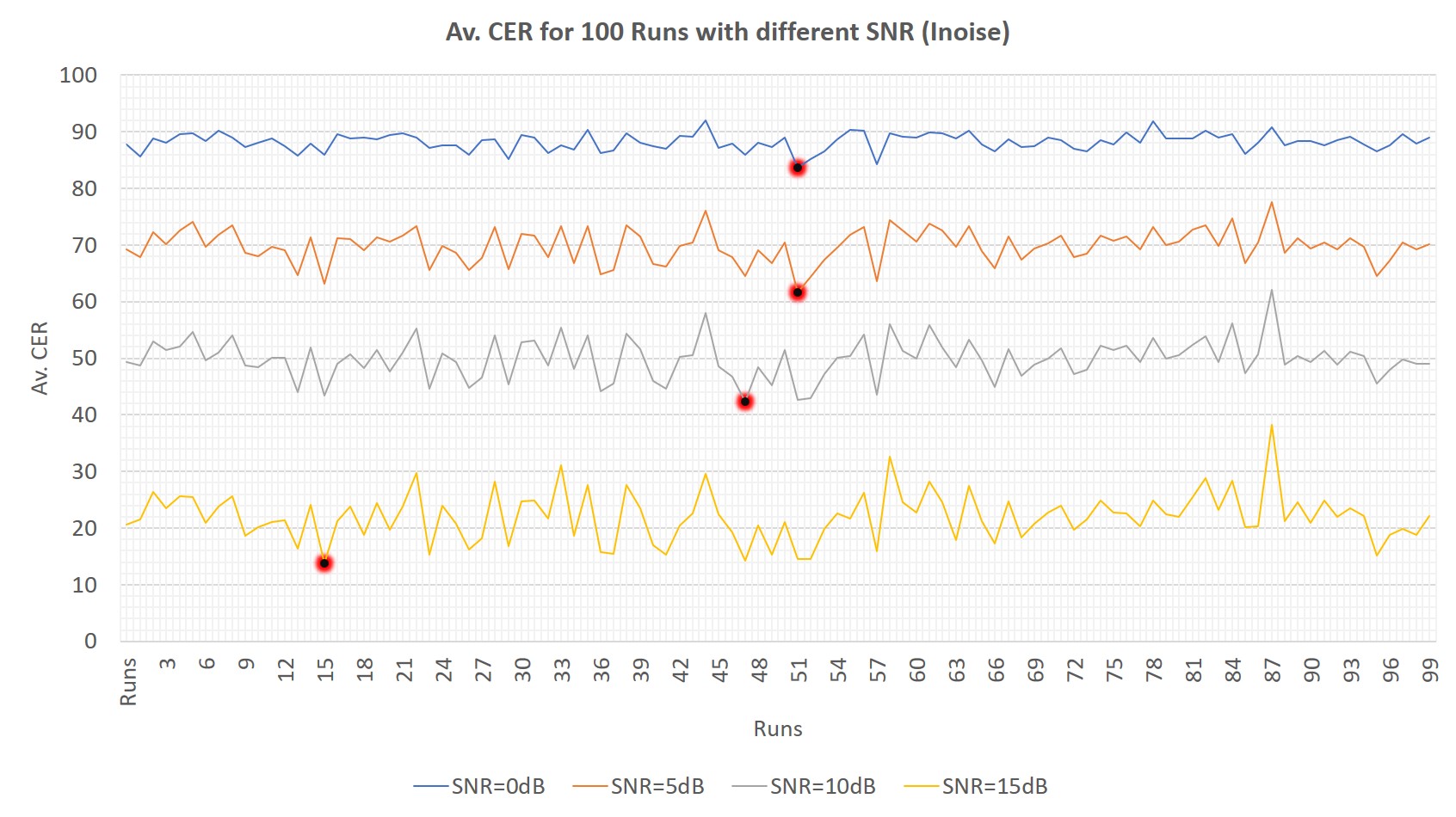}
    \caption{Average \CER\ for $100$ runs for varying SNR in iNoise (HOME) Noise.}
    \label{fig:inoise}
\end{figure}
\begin{table}
    \centering
    \caption{Optimal position of microphone $(x, y, z)$ for varying SNR signals. Note that the optimal position of the microphone is one that produces the minimum average \CER.}
    \begin{tabular}{|c||c|c|c|} \hline
    & \multicolumn{3}{c|}{Optimal microphone position} \\\cline{2-4}
SNR & $x_m^*$ & $y_m^*$ & $z_m^*$ \\ \hline
0 dB &1.0017  & 1.9992  & 2.9969 \\
5 dB &0.99825 &  2.00130  & 2.99950 \\
10 dB &0.99827 &  1.99910 &  2.99810 \\
15 dB &0.99765  & 1.99980 &  3.00170 \\ \hline
    \end{tabular}
    \label{tab:opt_mic_snr1}
\end{table}
As seen in Figure \ref{fig:noise_cer_runs1}, for Gaussian additive noise, the average \CER\ increases with higher amount of noise (i. e lower the SNRs). Note that the average \CER\ is at about $20$, $40$, $70$, $90$ for noise levels $15$, $10$, $5$, $0$ dB respectively. Also the number of Runs for which the average \CER\ is minimum are at $44$, $69$, $28$, $91$ for noise levels $15$, $10$, $5$, $0$ dB respectively. It is marked by a red circle in Figure \ref{fig:noise_cer_runs1}. Further, it may be noted that the optimal position (see Table \ref{tab:opt_mic_snr1}) of the microphone for different SNR is different.

We further used realistic noise to check the optimal position of the microphone. To perform this experiment, we took "HOME environment" noise from "iNoise database" \cite{w3xm-jn45-20}. The average \CER\ results are reported in Fig. \ref{fig:inoise}. 
 We found similar observations for {HOME} environment noise as well. With an increase in the SNR level, the average \CER\ decreases as observed for AWGN noise in Figure \ref{fig:noise_cer_runs1}. However, the average \CER  for AWGN noise is more or less flat for different runs unlike the realistic iNoise (HOME) conditions. Moreover, the average \CER\ is minimum at $52$, $52$, $48$, and $16$ number of runs at SNR $15$, $10$, $5$, and $0$ dB respectively and marked in red circle in Figure \ref{fig:inoise}. Note that the optimal position (see Table \ref{tab:opt_mic_snr2}) of the microphone for different SNR's is also different for iNoise (HOME) in comparison to the AWGN noise. It may again reiterated that the optimal position of the microphone is obtained that produces the minimum average \CER for each SNRs. 
 \begin{table}[!h]
    \centering
    \caption{Optimal position of microphone $(x, y, z)$ for varying SNR signals on iNoise (HOME) noisy environment.}
    \begin{tabular}{|c||c|c|c|} \hline
    & \multicolumn{3}{c|}{Optimal microphone position} \\\cline{2-4}
SNR & $x_m^*$ & $y_m^*$ & $z_m^*$ \\ \hline
0 dB &1.9713  & 2.3133  & 3.3930 \\
5 dB &1.9713 &  2.3133  & 3.3930 \\
10 dB &1.7906 &  2.7740 &  3.3386 \\
15 dB &1.1247  & 1.7584 &  2.7210 \\ \hline
    \end{tabular}
    \label{tab:opt_mic_snr2}
\end{table}

 This clearly illustrates the need of running Algorithm 1 separately for each noisy condition. This is due to the fact that the optimal position varies with the noise and SNR level. Further, it is demonstrating again that there is no easy way to identify the optimal location of the microphone analytically.

\section{Conclusions}
\label{sec:conclusions}
In this paper, we focused on studying if we could identify the true (or optimal) position of a microphone which might have been displaced in a single speaker, single microphone setup. We used Monte-Carlo simulation to randomly select $500$ possible microphone positions and then used each microphone location to generate RIR and the speech at that microphone location for $100$ different utterances. The generated speech was transcribed using a transformer based speech to alphabet converter (used in \cite{Nathwani2021} and \cite{Raikar2020}). For each of the microphone locations, we computed \cer\ and selected that microphone position which produced the minimum \cer\ (\ref{eq:mincer}) over all the $100$ utterances. The main contribution of the paper is in the experimental approach to identify the optimal location of the microphone. While we were keen on developing an analytical expression based on (\ref{eq:mincer}), it turns out that it might not be possible, given no strong correlation between the different microphone locations and the average \CER.  In summary, we identified the optimal location of the microphone, given the room size and speaker location. We used the average minimum \cer\ as the metric to find the best position of the microphone. We further showed that the optimal position of the microphone varies with the noise and SNR level of the signal as well.

\bibliographystyle{IEEEtran}
\bibliography{references}
\end{document}